\documentclass[10pt,superscriptaddress]{revtex4-2} 

\usepackage{soul}
\usepackage{mathtools}
\usepackage{amsmath}  
\usepackage{amsfonts} 
\usepackage{graphicx} 
\usepackage{xcolor}
\usepackage{mathrsfs}

\begin{document}


\title{Social Physics of Bacteria: \\ Avoidance of an Information Black Hole
}


\author{Trung V. Phan}
\thanks{These authors contributed equally to this work.}
\affiliation{Department of Chemical and Biomolecular Engineering, \\ John Hopkins University, Baltimore, MD 21218, USA.}

\author{Shengkai Li}
\thanks{These authors contributed equally to this work.}
\affiliation{Department of Physics, Princeton University, \\ Princeton, NJ 08544, USA.}

\author{Domenic Ferreris}
\affiliation{Department of Physics, Princeton University, \\ Princeton, NJ 08544, USA.}

\author{Ryan Morris}
\affiliation{School of Physics \& Astronomy, University of Edinburgh, \\ Edinburgh, EH9 3FD, Scotland, UK.}

\author{Julia Bos}
\affiliation{Unité Plasticité du Génome Bactérien, \\ Université Paris Cité, CNRS UMR 3525, Institut Pasteur, Paris, 75015, France.}

\author{Buming Guo}
\affiliation{Department of Physics, New York University, \\ New York, NY 10003, USA.}

\author{Stefano Martiniani}
\affiliation{Department of Physics, New York University, \\ New York, NY 10003, USA.}

\author{Paul Chaikin}
\affiliation{Department of Physics, New York University, \\ New York, NY 10003, USA.}

\author{Yannis G. Kevrekidis}
\affiliation{Department of Chemical and Biomolecular Engineering, \\ John Hopkins University, Baltimore, MD 21218, USA.}

\author{Robert H. Austin}
\affiliation{Department of Physics, Princeton University, \\ Princeton, NJ 08544, USA.}

\begin{abstract}
Social physics explores responses to information exchange in a social network \cite{social_physics}, and can be mapped down to bacterial collective signaling \cite{shapiro}. Here, we explore how social inter-bacterial communication includes coordination of response to communication loss, as opposed to solitary searching for food \cite{berg1975chemotaxis,keller1971model,trung},  with collective response emergence at the population level \cite{bonnie-ecoli,phan2020bacterial}. We present a 2-dimensional enclosed microfluidic environment \cite{morris2017bacterial} that utilizes concentric rings of funnel ratchets, which direct motile \textit{E.coli} bacteria towards a sole exit hole, an information ``black hole'', passage into the black hole irreversibly sweeps the bacteria away via hydrodynamic flow. We show that the spatiotemporal evolution of entropy production \cite{ro2022model} reveals how bacteria avoid crossing the hydrodynamic black hole information horizon.
\end{abstract}

\maketitle 

\section{Introduction}\label{sec1}

Bacterial chemotaxis can be both solitary \cite{berg1975chemotaxis} and communal \cite{bassler}.  The more familiar solitary form of chemotaxis, a lonely Proustian search of nutrients, can give rise to collective wave-like dynamics in a bacterial population. These observed collective motions emerge from individual chemotactic sensing and response to nutrient gradients created by other bacteria consuming their nutrients. As fascinating as these dynamics are, it can be difficult to attribute a definitive collective fitness advantage to these waves. However, bacteria can also communicate with each other directly in the absence of any metabolic gradients; this communication can be used to convey information related to the local time-dependent bacterial density $B(\vec r,t)$ \cite{quorum-review}. This collective sharing of information, and responding to this information, suggests an information approach to important aspects of bacterial behavior not considered in pure metabolic chemotaxis and which can be extended to other forms of active matter \cite{trevor}.

We confine our experiments and modeling to the bacteria {\em E. coli}, chemotactic bacteria which has an unusual quorum sensing response \cite{bonnie-ecoli}, our model discussed below explicitly takes this into account.  However, we believe the general phenomena and analysis presented here can be easily modified for other bacterial strains with differing sensing response. 

\section{The Information Black Hole Experiment: Technology}\label{sec2}

\begin{figure}[]
\centerline{\includegraphics[width=0.7\textwidth,keepaspectratio]{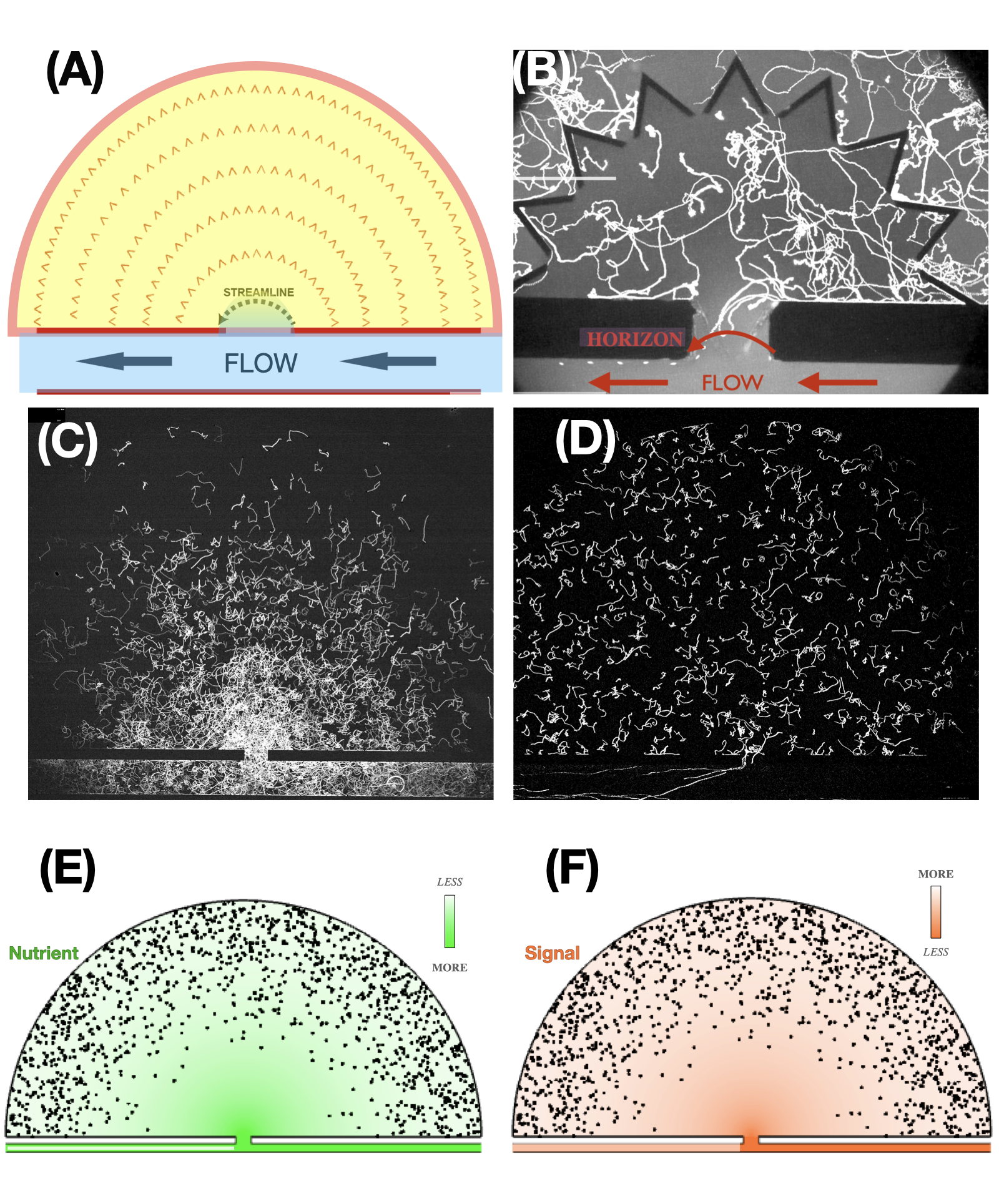}}
\caption{ (A) Schematic of the device. Nutrient rich flow is from right to left, the enclosed hemisphere has rings of funnels which direct motile bacteria towards the exit opening (the black hole). 
(B) Time lapse image of motile bacteria in an inward oriented funnel interior under no external flow conditions near the exit hole (black hole). One bacterium can be seen being swept away after crossing the flow separatrix. (C) Motile bacteria in a domain consisting of concentric funnel rings under no external flow condition at start of flow. (D) Motile bacteria in a round interior under no external flow conditions at start of flow. (E) Simulation of nutrient concentrations (brighter green denoted higher nutrient concentrations), and predicted bacterial densities in response to nutrient and signal. (F) Simulation of signal concentrations (brighter red denoted lower signal concentrations), and predicted bacterial densities in response to nutrient and signal.}
\label{fig1}
\end{figure}

We have constructed a microfabricated device which has both broken spatial symmetries and information sink-holes, allowing us to separate classical thermodynamic entropy from biological Shannon information entropy. The full device structure is depicted in the Supplementary Material, whilst a simplified diagram is shown in Fig. \ref{fig1}A. The exit hole is an opening into a channel in which the fluid flows by at an average speed $U$. This hydrodynamic black hole is the only exit from the enclosed volume: across this exit  there is no convective flow of fluid, but diffusion permits a finite flux of signaling molecules, media and bacteria across the streamline horizon. This exit hole, which we will call a \textit{black hole}, is bounded by a separatrix stream line; a bacterium crossing this streamline is swept away as shown in Fig. \ref{fig1}B in a recording of motile bacteria paths. The device shown in Fig. \ref{fig1}A has directed funnels (shown expanded in the photograph Fig. \ref{fig1}B); the funnels due to the placement of the open narrow apex of the funnel direct motile bacteria towards the exit hole.  In the absence of channel flow this results in a higher concentration of bacteria at the exit hole, as is seen in Fig. \ref{fig1}C. If we replace the funnels by round posts, so there is no symmetry breaking, the bacterial concentration remains roughly uniform in the device at least at the onset of flow, as seen in Fig. \ref{fig1}D.

The bacteria respond to two diffusing chemicals: local nutrients in the original nutrient rich media in the hemisphere and signaling molecules (the ``signal'') they exchange. Nutrients in the nutrient-rich flow diffuse across the exit hole into the hemisphere, where they are consumed as shown in Fig. \ref{fig1}E, whilst signaling molecules exchanged by the bacteria are carried away at the exit hole as shown in Fig. \ref{fig1}E as are  bacteria that enter it, as shown in Fig. \ref{fig1}F. As we will show, the net result between nutrient consumption and signal loss is what we will call an ``inverted'' bacterial population due to the conflicting gradients (nutrients vs signal) and loss of bacteria. To be more specific, the internal space in which the bacteria move was microfabricated into two different topologies:  (1) A ``flat'' space consisting of concentric half-circles of round posts, which biases motile bacteria neither towards nor away from the exit opening; and  (2)  a ``curved'' space where a bias like a form of gravity is induced by microfabricated concentric semicircles of triangular funnels, which channel motile (but not non-motile) bacteria to move in the direction of the narrow ends of the funnels \cite{galajda2007wall}. The net effect of funnels is to make the physical forward flux $J_F$ of the bacteria (``forward'' here means motion from  the large open end of the funnels towards their narrow opening) larger than the backward flux $J_B$ (``backward'' is simply the opposite direction of forward), thus violating time reversibility. The funnels physically bias the motion of motile bacteria so that $J_F > J_B$, which creates a net bacterial flux towards the hydrodynamic entry to the flow channel.  Fig. \ref{fig1}E and F, taken from the simulations described below, shows the inverted populations created when lack of signal molecules, due to their exiting through the hydrodynamic black holes, overwhelms nutrient chemotaxis, as described below.

In our experiments we used a modified form of the common laboratory {\em E. coli} strain W3110. We used the strain (RP437 att::gfp), which continuously expresses GFP in the cell body for accurate tracking of the bacteria under the microscope \cite{delgado2021efficient}. 

\section{Information Black Hole: \\ Streamline border, Bacterial and Signal Return Probabilities}\label{sec8}

The Reynolds number $Re$, which quantifies the ratio between inertial and viscous forces of the flow in our device, is always much less than $1$.  To see this, we assume
$Re \sim \frac{\rho U \Delta}{\eta}$, where $\rho$ is the density of water, $U$ is the fluid velocity, $\Delta$ is a characteristic turning distance for the fluid flow and hence the size of the exit hole and $\eta$ is water viscosity.  Using $\Delta \sim$ 10$^2 \mu$m and $U \sim$ 10$^2\mu$m/s, we get $R_e \sim$ 10$^{-1}$, so the flow is always laminar and not turbulent.  

The P\'{e}clet number Pe of the black hole determines if re-entry of an object which crosses the streamline horizon is unlikely.  If the black hole has a diameter $L$, and the object has a diffusion coefficient $D$, then the Peclet number is defined as the ratio of the time $t_D$ to diffuse the distance $L$ and the time $t_A$ to advect the same distance, given that the fluid has a speed $U$ \cite{peclet}: $
 Pe = \frac{L^2/D}{L/U} = \frac{LU}{D}$
where $D \sim \frac{k_BT}{6 \pi \eta r}$, where $T$ is the absolute temperature, $\eta$ is the solvent viscosity and $r$ is the characteristic size of the object. If $Pe \gg 1$ then advection dominates and horizon passage is irreversible. For bacteria in water, $D_{bac} \sim$10$^{-9}$ cm$^2$/s, while a typical peptide molecule used in quorum signaling \cite{quorum-peptide} has a $D_{s} \sim $10$^{-6}$ cm$^2$/s.  Since our black hole has a length $L$ of 10$^{2}$ $\mu$m, and our flow velocities U are on the order of 10$^{-2}$ cm/s, the bacteria have a $Pe_{bac} \sim$10$^2$, while the signaling molecules have a $Pe_s \sim$10$^{-1}$. We calculate that the flow has a $Re \ll 1$ and the bacteria have a $Pe_{bac} \gg 1$. This means, bacteria which cross a streamline across the face of the black hole due to motility are irreversibly swept way, while signaling molecules will have a roughly 50\% loss rate, and nutrient molecules can with good probability enter the enclosed volume from the flow channel.

\section{Experimental Results}

We did experiments both with bacteria in the presence of rich LB broth were we expect strong nutrient chemotaxis (and bacterial reproduction) to occur, and in minimal media media, which contains only  glucose as a carbon source and essential salts. In the minimal media case, we would expect that nutrient chemotaxis to be minimal since of the 5 known chemoreceptors in {\em E. coli} (Tar, Tsr, Tap, Trg, Aer) only the Trg receptor would be activated \cite{5-receptors}, and the spatial population dynamics to be governed primarily by hydrodynamic pumping of motile bacteria and information-based collective response. 

\subsection{No-flow: Pure Ratchet Sorting}

If our bacteria move randomly, the probability $p_i$ to find them in a particular pixel as a function of position $\vec{r}_i$ would be a constant when averaged over time. In the case of a concentric ring of funnels, there is a physics-based symmetry-breaking bias in overall bacterial densities towards the central exit hole when there is no net flow in the channel. In the case of funnels this creates a radially dependent bacterial density $B(r)$ for motile bacteria, which peaks at the exit hole entrance in the absence of flow, or in the case of replacing the funnels by round posts a uniform distribution of motile bacteria within the volume before flow is applied. Fig. \ref{fig1}C shows the trajectories of motile bacteria for a funnel interior, while in Fig. \ref{fig1}D we plot trajectories of motile bacteria in an interior where there are no funnels, but cylindrical posts instead. As expected, motile bacteria are pumped by the array of funnels into the hole under no flow conditions, while for posts no such pumping for motile bacteria occurs and the bacteria are uniformly dispersed, as long as the bacterial population is not so high as to allow nutrient gradients to develop, and low enough that inter-bacterial signaling is not important.

\begin{figure}[ht]
\centering
\includegraphics[width=0.7\textwidth,keepaspectratio]{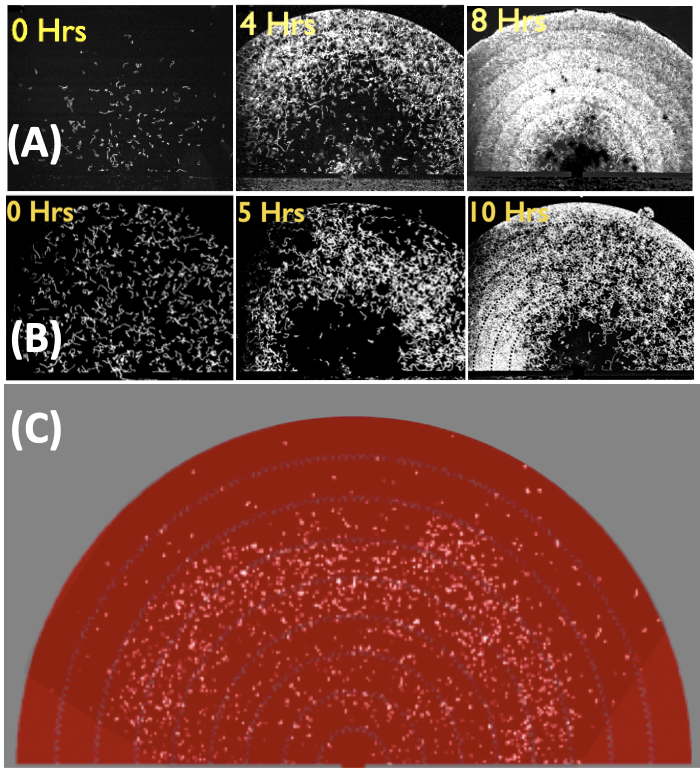}
\caption{\textbf{Development of Inverted Bacterial Populations.} (A) Development of an inverted population of bacteria in an inward directed funnel (``gravity'') environment in the presence of nutrient rich LB. (B) Development of an inverted population of bacteria in an flat (round posts) environment in the presence of nutrient rich LB. (C) Inverted population of bacteria in minimal media after 2 hours of flow. We have superposed the funnel array to guide the eye. Camera framing cut off the image in the corners.}
\label{fig2}
\end{figure}

\subsection{Development of Inverted Bacterial Populations}

Once flow is turned on, the external hydrodynamic flow sweeps away bacteria which cross the stream-line, and to a lesser extent signaling molecules are also swept away. Bacteria away from the exit hole will then see a gradient in the signaling molecules develop, which points away from the black hole, even as a nutrient gradient develops towards the exit hole, while the funnel ring continues to transport motile bacteria towards the exit hole. Only motile bacteria are imaged here. Adhesion of bacteria to the 
silicon dioxide surface of silicon resulted in bound non-motile bacteria, their fluorescent images were subtracted from the images. Only bacteria which changed position from frame to frame are shown. 

Figure \ref{fig2}A shows the development of an inverted population of bacteria in an inward directed funnel (``gravity'') environment in the presence of nutrient rich LB. Due to the rich media, the bacteria divide during the experiment and show rapid growth in bacterial density away from the black hole.  However, in spite of the presence of un-metabolized media within the flow channel diffusing into the hemisphere, creating a strong chemotactic gradient, and the presence if funnels which drive motile bacteria towards the black hole, a very clear unoccupied region develops around the black hole, into which bacteria do not enter. 

We also explored a ``flat'' landscape, that is, one where there are simply round posts that provide no bias in (gravity) from the black hole.  Fig. \ref{fig2}B shows that in this case we start with a flat population and smoothly develop into an inverted population.

Finally, we probed what happens when the media is not rich, that is, only contains glucose as a sole carbon source. In that case only the Trg receptor would be activated, and we would expect that nutrient chemotaxis would be greatly inhibited.  These turned out to be difficult experiments due to the very slow growth of the bacteria in minimal media and the attendant low production of gfp as a marker, making for a very dim field with low densities of bacteria. Nevertheless as we show in Fig. \ref{fig2}C the same phenomena of population inversion and black hole avoidance still occurred, indicating the generality of the phenomena.

\section{Modeling Bacterial Communication}

We present two ways to view this phenomena of population inversion, both involving information transfer between bacteria.  Keller and Segel \cite{keller1971model} developed a set of partial differential equation (KSE) to describe the chemotaxis of bacteria along nutrient gradients and the growth and death of bacteria, but with no inter-bacterial information exchange, hence two fields are used: bacterial density $B(\vec{r},t)$ and the nutrient density $N(\vec{r},t)$. A third field, an information field $A(\vec{r},t)$ is needed to complement the KSE if bacteria signal directly to each other in the absence of metabolites. Brenner et al \cite{brenner1998physical} introduced a field of signaling molecules $A$, but did not include nutrient chemotaxis (the BKSE model) and relied on nutrients to generate a signal. In these models we do not take into account cell phenotype heterogeneity, which results in interesting selection dynamics \cite{fu2018spatial, mattingly2022collective}. 

\begin{figure}[ht]
\centering
\includegraphics[width=1.0\textwidth,keepaspectratio]{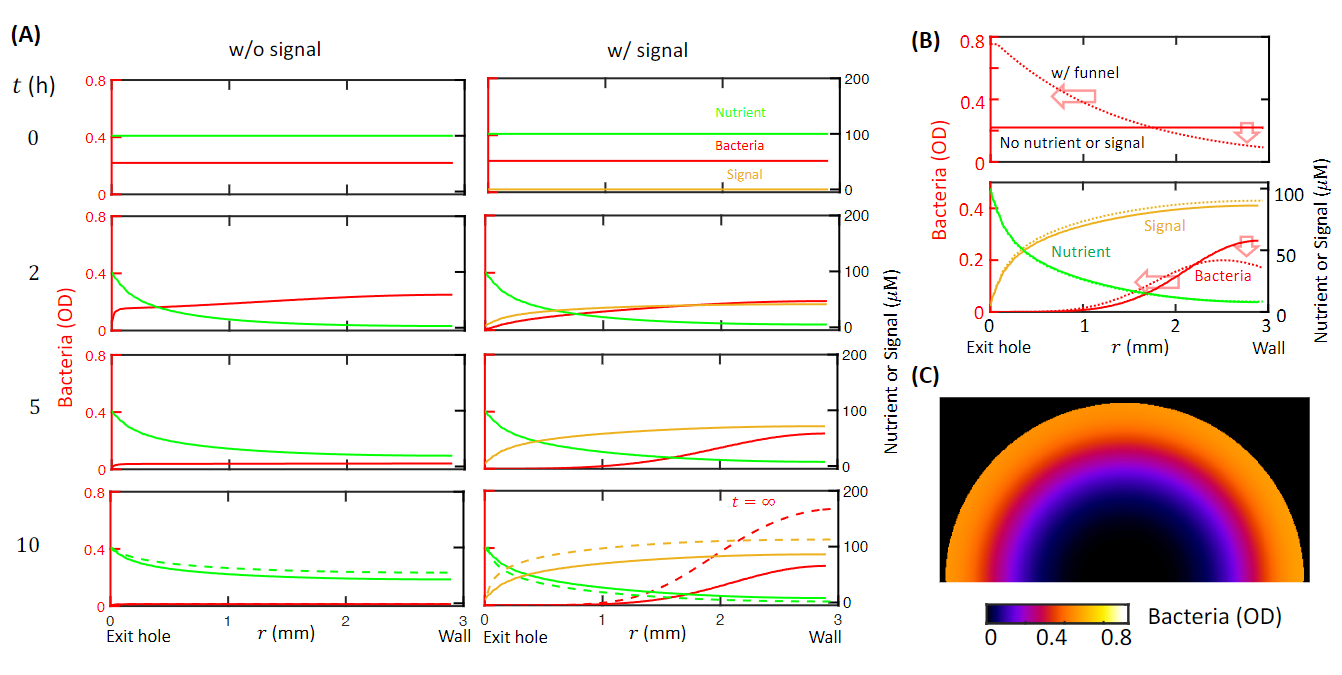}
\caption{\textbf{Signal induced inverted population.} (A) Left: time evolution of the bacteria and nutrients following the original KSE without signal. Right: time evolution of the bacteria, nutrients, and the signal following NSE with $\kappa_A/\kappa_F=3$, in which the bacteria chemotax along the signal secreted by themselves. The dash lines in the $t=10$ h panels show the field at $t=\infty$ (steady state). The curves are all $\theta=90^\circ$ cuts of the full 2D PDE integration with $t=10$ h snapshots shown in the bottom row. (B) Inverted population persists against the ``gravity'' created by funnels: As shown in (B), when a funnel with a linear effective potential ($\Phi_F=C_F-k_F|\mathbf{r}|$ where $k_F=0.33~\mu$m/s. See Sec.\ref{sec:funnel}) is installed, the bacteria (dotted lines) is more concentrated in the funnel direction (towards left) compared with the case without the funnel, as shown in (A). The boundary conditions for (A) and (B) are no-flux (Neumann) at both ends. When the linear funnel is installed in the NSE equations, compared to the result without funnels (solid lines), which is shown in (C), a reproduction of Fig.\ref{fig:PDE_evol} at $t=10$ h, the inverted population is pulled a little bit along the funnel and towards the black hole, but the inverted population as shown in (D) persists.}
\label{fig:PDE_evol}
\end{figure}

Clearly, real bacteria both sense nutrients (they must find food to survive), and they sense the presence of other bacteria and respond to their presence.  We also include in our model the fact that {\em E. coli} at high densities do not produce the signaling molecule (A) \cite{bonnie-ecoli}.
Thus, we have modified the BKSE to include both nutrient and bacterial signals, and made the signal production ratio exponentially decreasing with higher bacteria concentrations:
\begin{eqnarray}
\partial_t B &=& \vec{\nabla}^2 (D_B B) + \alpha N \left(1-\frac{B}{B_s} \right)B \nonumber \\
& & - \vec{\nabla} \cdot \left( B  \left\{ \kappa_A \vec{\nabla} \Phi_A[A] + \kappa_N \vec{\nabla} \Phi_N[N] \right\} \right) \ , \label{eq:BKS_B} \\
\partial_t A &=& \vec{\nabla}^2 (D_A A) + \beta B ~ e^{-B/B_A} \ , \label{eq:BKS_A}\\
\partial_t N &=& \vec{\nabla}^2 (D_N N) - \gamma BN \ ,
\label{Shengkai}
\end{eqnarray}
where $B(\vec{r},t)$, $N(\vec{r},t)$, and $A(\vec{r},t)$ are the bacterial density, nutrient and signal concentrations, respectively. The functions $\Phi_A[A]$ and $\Phi_N[N]$ are the perceived strength of the nutrient and signal concentrations as felt by the bacteria \cite{trung}; for our model we use a linear relationship: $\Phi_A[A]=A$ and $\Phi_N[N]=N$. Equations \ref{Shengkai} we call NSE, for Nutrient+ Signal+Equations. They deserve a new name because in addition to the terms in classical KSE, including the consumption of nutrients by the bacteria ($-\gamma BN$), the logistic growth of bacteria ($+\alpha NB(1-B/B_s)$) in which $B_s$ represents the carrying capacity, and the diffusion of bacteria and nutrients, we mix the signaling field $A$ secreted by the bacteria with the nutrient field $N$. Further, we assume the signaling molecules secreted per bacteria decrease at high bacteria bacterial densities ($+\beta B ~ e^{-B/B_A}$), a form of quorum sensing which prevents total spatial population collapse into a singularity \cite{bonnie-ecoli}.

The boundary at the exit hole represents an event horizon: the hydrodynamic flow is so fast that the resource is constantly replenished to a high value but entering bacteria crossing a critical streamline  are flushed away. If we consider the small but finite flow of bacteria out into the fast-flowing channel, Robin (radiation) boundary conditions \cite{robin} are essential in solving the NSE equations. In Robin boundary conditions, the concentration of a substance $u$ (nutrients, signals, or bacteria) is a function of internal and external concentrations:
\begin{eqnarray}
\vec{\nabla}_{\hat{n}} u = K (u-u_0) \label{eq:RBC}
\end{eqnarray}
where $\hat{n}$ is the direction perpendicular to the spatial boundary and $u_0$ is the constant concentration outside. In our system, the outside concentration for bacteria and signal is $0$. The outside concentration for nutrients is $N_0$. $K$ sets the strength of the radiation of substance. See II.B of the Supplementary Material for details.

The arena diameter is 6000 microns and the opening of black hole is  200 microns. We set the effective diffusion coefficient of a motile  bacterium $D_B=200~\mu$m$^2/$s. The diffusion coefficient of the signal molecules ($D_A$) and metabolites ($D_N$) was set to 800 $\mu$m$^2/$s \cite{trung}. The bacterial reproduction rate $\alpha_0$ was set  to $0.5/$h in rich media ($\alpha=\alpha_0/N^*$ where $N^*=100~\mu$M). The chemotaxis sensitivity for nutrients  $\kappa_N$ was set to be the same value used for aspartate chemotaxis,  2400 $\mu$m$^2$/$\mu$M$~$s \cite{trung}, and we let the signaling sensitivity $\kappa_A$ be a variable. The inverted population emerges (Fig.\ref{fig:PDE_evol}) when $\kappa_A \gtrsim \kappa_N$.

Fig. \ref{fig:PDE_evol}(A) left shows that in the case with only nutrient chemotaxis (the KSE model), the chemotaxis gradient constantly points towards the black hole, leading eventually to a complete wash-out of the bacteria in search of food. However, Fig. \ref{fig:PDE_evol}(A) Right shows that when the bacteria sense an attractive signaling field  produced by the bacteria itself, as occurs for example in quorum sensing, the bacteria distribution evolves into an inverted population, away from the exit hole (black hole), due to loss of signal.

Intuitively, when there are signaling molecules, as the bacteria disappear into the exit hole, the signaling molecule concentration decreases, in essence a warning for the other bacteria to stay away. The gradient from this dip around the exit hole makes the bacteria aggregate at some place away from the  exit hole, which represents an existential loss of fellow bacteria.

\subsection{Inverted population persists even in inward funneling}\label{sec:funnel}

We used an an effective potential function $\Phi_F(\vec{r})$ whose spatial gradient produces the pumping action of funnels on motile bacteria. Funnel pumping action on motile bacteria violates time-reversal symmetry \cite{ro2022model} so the local funnel field should have local non-zero curl at the funnels, but for ease of computation we have replaced the local curl derived pumping action with a smoothed global force gradient. We modify the bacterial directional response in Eq. \eqref{eq:BKS_B}, as follows:
\begin{eqnarray}
\kappa_A \vec{\nabla} \Phi_A[A] + \kappa_N \vec{\nabla} \Phi_N[N] \ \ \longrightarrow \ \ \kappa_A \vec{\nabla} \Phi_A[A] + \kappa_N \vec{\nabla} \Phi_N[N] + \vec{\nabla} \Phi_F(\vec{r}) \ .
\end{eqnarray}
For simplicity, we consider a radially linear dependence for $\Phi_F(\vec{r})=C_F-k_F\left|\vec{r}\right|$. The gradient of  $\Phi_F$ is set to zero at the wall boundaries ($\partial_r \Phi_F = 0$, see the appendix for details) to conserve the total mass of bacteria.

\section{Entropy Production via Communication}

\begin{figure}[ht]
\centering
\includegraphics[width=1.0\textwidth,keepaspectratio]{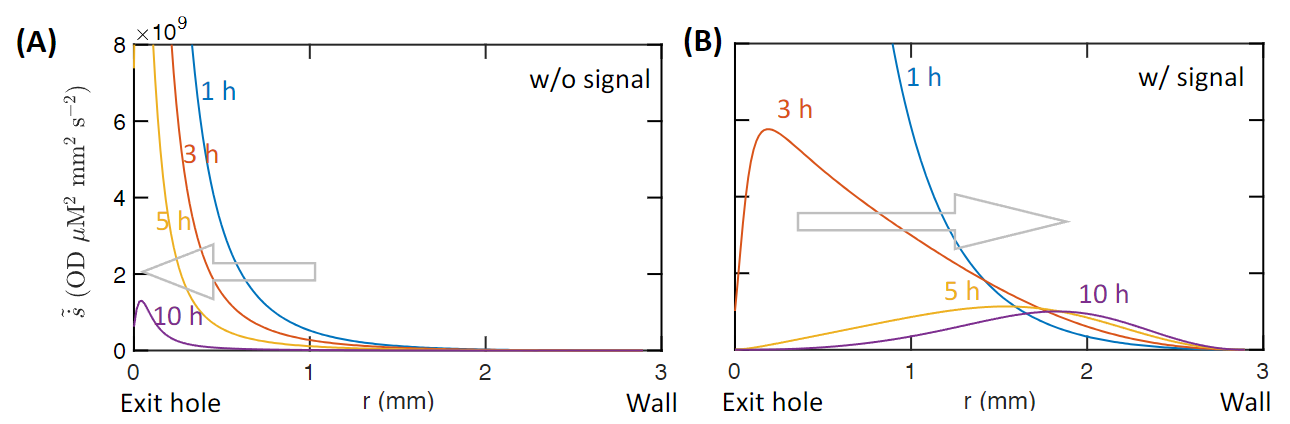}
\caption{\textbf{Signal Entropy Production} The normalized entropy production rates for bacteria sensing nutrient and for bacteria sensing both nutrient and signal are shown in (A) and (B) respectively. Here the ``reduced'' entropy production rate is $\tilde{\dot{s}}=B~(\kappa_N\nabla N+\kappa_A\nabla A)^2$, which is proportional to the lower bound shown in Eq.\ref{eq:entropyProd}.}
\label{fig:entropy}
\end{figure}

Entropy production by motile bacteria has been studied regarding the time-irreversible bacterial flow induced from their mechanical interactions with the funnel ratchets \cite{ro2022model}. However, time-irreversible bias in the collective migration of bacteria can also emerge due to the flow of communication information (via signaling molecules and nutrients distribution), generating another contribution to entropy production \cite{nguyen2023remark}. Let us call the former the \textit{mechanical contribution} and the later the \textit{communication contribution}. Hagen et al \cite{quorum-review} pioneered this kind of analysis from a physical viewpoint, and we expand on this collection of works with quorum sensing.

The theoretical foundation of entropy production stems from stochastic thermodynamics, in which the derivative of the Shannon entropy (which we denote as $\dot{S}$) gives rise to two important quantities: the entropy production rate and the entropy flow rate \cite{van2015ensemble}. The non-negative entropy production rate $\dot{S}_i \geq 0$ quantifies the time-irreversibility and dissipative nature of the system, and the entropy flow rate $\dot{S}_f$ quantifies the net flow of entropy. In combination, $\dot{S}_i + \dot{S}_f = \dot{S}$. Even when our system reaches a steady-state and the Shannon entropy stays constant $\dot{S}=0$, the entropy production rate and the entropy flow rate can still be non-zero i.e. $\dot{S}_i = -\dot{S}_f \neq 0$, indicating that there are still entropy fluxes to keep the system at a configuration far away from thermal equilibrium, due to information transfer. Besides this informatic interpretation, studying the entropy production rate $\dot{S}_i$ is also crucial, as it helps estimate the maximum locally extractable mechanical power from the system, which is an important task in bioengineering \cite{ro2022model,nguyen2023remark}.

Since the \textit{mechanical contribution} to entropy production has been well-understood in previous work \cite{ro2022model}, here we consider the \textit{communication contribution}. The entropy production rate is a quantification of the time-irreversibility of local signaling dynamics, which can be estimated by the non-negative Kullback-Leibler divergence \cite{kullback1951information} between the probability distribution of all possible local state trajectories $\chi$ during a time interval $\tau$ and that of their time-reversal realizations $\chi^R$:
\begin{equation}
\dot{S}_p = \tau^{-1} \left\langle \ln \frac{\mathcal{P}[\chi]}{\mathcal{P}[\chi^{R}]} \right\rangle \geq 0 \ .
\label{local_EP}
\end{equation}
Here, the calculation of Shannon entropy involves associating microstates with the spatial occupancy of bacteria; therefore, each trajectory $\chi$ represents the history of transitions between states. A derivation for the communication entropy production rate, starting from the adaptive run-and-tumble chemotatic response of \textit{E.coli} cells, can be found in \cite{nguyen2023remark}. This estimation is based on field gradients, which directly influence the local bias of bacterial trajectories: 
\begin{equation}
 \dot{S}_i \geq \left( \frac{3 k_{\text{b}} }{v_0} \right) B \left\{ \kappa_A \vec{\nabla}\Phi_A\left[A\right] + \kappa_N \vec{\nabla}\Phi_N\left[N\right] \right\}^2 \ ,
 \label{eq:entropyProd}
\end{equation}
where $k_b$ is the Boltzmann constant and $v_0 \sim 30\mu$m/s is the swimming speed of \textit{E.coli}. Eq. \ref{eq:entropyProd} was used in Fig. \ref{fig:entropy} in conjunction with Eq. \ref{Shengkai} to graphically show the spatial dependence of signaling entropy production.  However, since we cannot image the actual signaling molecule fields, at present this cannot be used as a quantitive proof of the model. In any event, the gradients of chemical fields (signaling molecules and nutrients) determine the entropy production in communication, and their scaling-proportionality (to the square-power) is consistent with the measured information acquisition rate of a single bacterium during generalized chemotaxis \cite{mattingly2021escherichia}.

\section{Conclusions}

Neither the classical nutrient-only Keller-Segel equations, nor the modified Brenner-Keller-Segel equations, which utilize nutrient derived signaling can fully capture the experimental response of bacteria to an environment where loss of signal trumps nutrient gradients. Signaling and loss of signal in a localized region of space, our exit black hole, results in the collective movement of the bacteria away from the silent region in spite of high nutrients in the black hole. The social physics of {\em E. coli} population dynamics must include both an individualist search for food and the necessity of finding other bacteria, with the surprising experimental result that increased communication within a high population density easily trumps lack of food in that region. An information theoretical analysis provides insights into the far-from equilibrium dynamics the system ehibits. Thus bacteria can collectively perform complex tasks critical for the population's survival using information exchange, including evading existential threats such as the loss of information as observed in our experiments.

\vskip0.1in
\section{Acknowledgements}

This work was supported by the US National Science Foundation (PHY-1659940 and PHY-1734030).

\bibliography{main}

\begin{thebibliography}{10}

\bibitem{social_physics}
E.~Ben-Jacob.
\newblock Social behavior of bacteria: from physics to complex organization.
\newblock {\em European Physical Journal B}, 65(3):315--322, 2008.

\bibitem{shapiro}
J.~A. Shapiro.
\newblock The significances of bacterial colony patterns.
\newblock {\em Bioessays}, 17(7):597--607, 1995.

\bibitem{berg1975chemotaxis}
Howard~C Berg.
\newblock Chemotaxis in bacteria.
\newblock {\em Annual review of biophysics and bioengineering}, 4(1):119--136, 1975.

\bibitem{keller1971model}
Evelyn~F Keller and Lee~A Segel.
\newblock Model for chemotaxis.
\newblock {\em Journal of theoretical biology}, 30(2):225--234, 1971.

\bibitem{trung}
Trung~V Phan, Henry~H Mattingly, Lam Vo, Jonathan~S Marvin, Loren~L Looger, and Thierry Emonet.
\newblock Direct measurement of dynamic attractant gradients reveals breakdown of the patlak--keller--segel chemotaxis model.
\newblock {\em Proceedings of the National Academy of Sciences}, 121(3):e2309251121, 2024.

\bibitem{bonnie-ecoli}
M.~G. Surette and B.~L. Bassler.
\newblock Quorum sensing in e. coli and s. typhimurium.
\newblock {\em Proceedings of the National Academy of Sciences of the United States of America}, 95(12):7046--7050, 1998.

\bibitem{phan2020bacterial}
Trung~V Phan, Ryan Morris, Matthew~E Black, Tuan~K Do, Ke-Chih Lin, Krisztina Nagy, James~C Sturm, Julia Bos, and Robert~H Austin.
\newblock Bacterial route finding and collective escape in mazes and fractals.
\newblock {\em Physical Review X}, 10(3):031017, 2020.

\bibitem{morris2017bacterial}
Ryan~J Morris, Trung~V Phan, Matthew Black, Ke-Chih Lin, Ioannis~G Kevrekidis, Julia~A Bos, and Robert~H Austin.
\newblock Bacterial population solitary waves can defeat rings of funnels.
\newblock {\em New Journal of Physics}, 19(3):035002, 2017.

\bibitem{ro2022model}
Sunghan Ro, Buming Guo, Aaron Shih, Trung~V Phan, Robert~H Austin, Dov Levine, Paul~M Chaikin, and Stefano Martiniani.
\newblock Model-free measurement of local entropy production and extractable work in active matter.
\newblock {\em Physical review letters}, 129(22):220601, 2022.

\bibitem{bassler}
Melissa~B Miller and Bonnie~L Bassler.
\newblock Quorum sensing in bacteria.
\newblock {\em Annual Reviews in Microbiology}, 55(1):165--199, 2001.

\bibitem{quorum-review}
Stephen Hagen.
\newblock {\em The Physical Basis of Bacterial Quorum Communication}.
\newblock Springer Verlag, New York, NY 10036, USA, 2015.

\bibitem{trevor}
T.~GrandPre, K.~Klymko, K.~K. Mandadapu, and D.~T. Limmer.
\newblock Entropy production fluctuations encode collective behavior in active matter.
\newblock {\em Physical Review E}, 103(1), 2021.

\bibitem{galajda2007wall}
Peter Galajda, Juan Keymer, Paul Chaikin, and Robert Austin.
\newblock A wall of funnels concentrates swimming bacteria.
\newblock {\em Journal of bacteriology}, 189(23):8704--8707, 2007.

\bibitem{delgado2021efficient}
Josemar{\'\i}a Delgado-Mart{\'\i}n and Leonardo Velasco.
\newblock An efficient dsrna constitutive expression system in escherichia coli.
\newblock {\em Applied Microbiology and Biotechnology}, 105(16-17):6381--6393, 2021.

\bibitem{peclet}
Vogel S.
\newblock {\em Life in Moving Fluids. The Physical Biology of Flow}.
\newblock Princeton University Press, Princeton, NJ, 08544, USA, 1996.

\bibitem{quorum-peptide}
F.~Verbeke, S.~De~Craemer, N.~Debunne, Y.~Janssens, E.~Wynendaele, C.~Van~de Wiele, and B.~De~Spiegeleer.
\newblock Peptides as quorum sensing molecules: Measurement techniques and obtained levels in vitro and in vivo.
\newblock {\em Frontiers in Neuroscience}, 11:1--18, 2017.

\bibitem{5-receptors}
S.~Weerasuriya, B.~M. Schneider, and M.~D. Manson.
\newblock Chimeric chemoreceptors in escherichia coli: Signaling properties of tar-tap and tap-tar hybrids.
\newblock {\em Journal of Bacteriology}, 180(4):914--920, 1998.

\bibitem{brenner1998physical}
Michael~P Brenner, Leonid~S Levitov, and Elena~O Budrene.
\newblock Physical mechanisms for chemotactic pattern formation by bacteria.
\newblock {\em Biophysical journal}, 74(4):1677--1693, 1998.

\bibitem{fu2018spatial}
Xiongfei Fu, Setsu Kato, Junjiajia Long, Henry~H Mattingly, Caiyun He, Dervis~Can Vural, Steven~W Zucker, and Thierry Emonet.
\newblock Spatial self-organization resolves conflicts between individuality and collective migration.
\newblock {\em Nature communications}, 9(1):2177, 2018.

\bibitem{mattingly2022collective}
Henry~H Mattingly and Thierry Emonet.
\newblock Collective behavior and nongenetic inheritance allow bacterial populations to adapt to changing environments.
\newblock {\em Proceedings of the National Academy of Sciences}, 119(26):e2117377119, 2022.

\bibitem{robin}
K.~Gustafson and T.~Abe.
\newblock The third boundary condition—was it robin’s?
\newblock {\em The Mathematical Intelligencer}, 20:63–71, 1998.

\bibitem{nguyen2023remark}
Minh~DN Nguyen, Phuc~H Pham, Khang~V Ngo, Van~H Do, Shengkai Li, and Trung~V Phan.
\newblock Remark on the entropy production of adaptive run-and-tumble chemotaxis.
\newblock {\em Physica A: Statistical Mechanics and its Applications}, page 129452, 2023.

\bibitem{van2015ensemble}
Christian Van~den Broeck and Massimiliano Esposito.
\newblock Ensemble and trajectory thermodynamics: A brief introduction.
\newblock {\em Physica A: Statistical Mechanics and its Applications}, 418:6--16, 2015.

\bibitem{kullback1951information}
Solomon Kullback and Richard~A Leibler.
\newblock On information and sufficiency.
\newblock {\em The annals of mathematical statistics}, 22(1):79--86, 1951.

\bibitem{mattingly2021escherichia}
HH~Mattingly, K~Kamino, BB~Machta, and T~Emonet.
\newblock Escherichia coli chemotaxis is information limited.
\newblock {\em Nature physics}, 17(12):1426--1431, 2021.

\end{thebibliography}
\bibliographystyle{unsrt}

\end{document}